\documentclass{llncs}

\usepackage{graphics}
\usepackage{graphicx}
\usepackage{url}
\usepackage{multirow}
\usepackage[font=small,labelfont=bf]{caption}

\usepackage{url}
\title{{\bf A Common Evaluation Setting for Just.Ask, Open Ephyra and Aranea QA systems}}
\author{{\normalsize Ricardo Pires}}
\institute{{\normalsize IST - Instituto Superior T\'ecnico}\\
        {\normalsize {\tt rmpp@ist.utl.pt}}}
\date{}

\begin{document}
\maketitle
\thispagestyle{empty}
\begin{abstract}
Question Answering (QA) is not a new research field in Natural Language Processing (NLP). However in recent years, QA has been a subject of growing study. Nowadays, most of the QA systems have a similar pipelined architecture and each system use a set of unique techniques to accomplish its state of the art results. However, many things are not clear in the QA processing. It is not clear the extend of the impact of tasks performed in earlier stages in following stages of the pipelining process. It is not clear, if techniques used in a QA system can be used in another QA system to improve its results. And finally, it is not clear in what setting should be these systems tested in order to properly analyze their results.

\end{abstract}

\section{Introduction}

Evaluating complex systems like QA systems is a very difficult task. Although QA systems have a similar architecture they use very different methods to accomplish their goals. As a consequence, it is hard to have a good evaluation setting for these different QA systems. In the last decade several competitions (TREC and CLEF for example) have been organized year after year to test different systems on the same task. These competitions only evaluate QA systems extrinsically, e.g., by its overall results. Although these informations are useful, they cannot be used to evaluate the relevance of the criteria that the different systems use, neither the contribution of each component. 
The main goal of this work is to establish a evaluation setting where three QA systems -– Just.Ask, Aranea and Open Ephyra - can be properly evaluated both extrinsically and intrinsically. Regarding the elaboration of this evaluation setting is important to know: what types of questions should be used; how do the systems process their input; what are the important characteristics that the corpora must possess; what evaluation measures should be used etc. A depth understanding of the systems at hand is needed in order to recognize the relevant characteristics for the evaluation process. 

The organization of this work is as follows: first, Section 2 gives a little explanation about the terminology that will be used throughout the work. The description of Just.Ask, Aranea and Open Ephyra architectures will presented in Sections 3, 4, and 5 respectively. Section 6 presents the preliminary evaluation of these systems which identifies the most important characteristics to take into account for a proper future evaluation of these systems. The rationale for the choose of the best evaluation setting is given in Section 7. Section 8 describes some problems that occurred in adapting the systems to this setting. Sections 9, 10 and 11 describe respectively, the evaluation of Aranea and extrinsic and intrinsic evaluation of Open-Ephyra and Just.Ask. Section 12 discusses the feasibility of improvements discovered throughout this work and finally Section 13 summarizes the topics discussed in this work.

\section {Standard QA Architecture} 

The generalized architecture of a QA system has become standardized \cite{jurafsky08}, consisting of a pipeline with 3 processing stages: Question Interpretation, Passage Retrieval and Answer Extraction. Question Interpretation module receives the user question and has two main goals: to formulate queries based in information extracted from the question - Query Formulation - and to classify the question by its expected answer type - Question Classification.Passage Retrieval module The formulated queries are next used    
 In the following sections figures that display the architectures of Just.Ask, Aranea and Open Ephyra systems may have different names for these three processing stages of the standard QA pipeline. In order to use the same terminology throughout this work, the standard QA pipeline terminology will be used in describing those architectures. 

\section {Just.Ask Architecture}
 
Just.Ask is a traditional ontology-driven system that utilizes the standard QA pipelined architecture (as showed in Figure 1).

\begin{figure}[h!t!]
\centering
\includegraphics[scale=0.6]{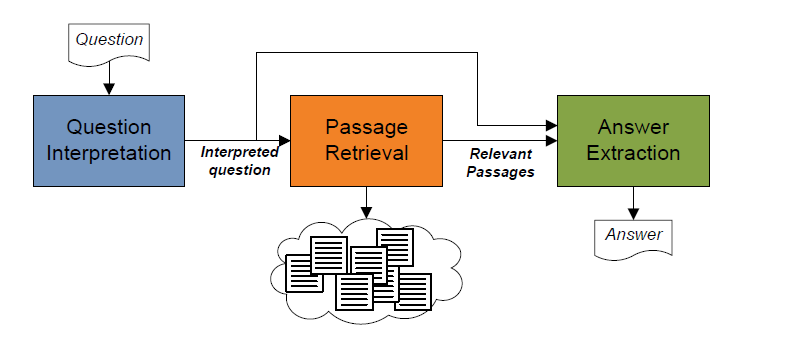}
\caption{Just.Ask Architecture}
\label{fig:QAPipelining}
\end{figure}

\subsection {Question Interpretation}

The Question Interpretation module is responsible for understanding the question. It receives as input a question from the user, expressed in natural language and outputs an interpreted question.  The interpreted question includes the original user question and useful information for the Passage Retrieval module. This module performs two main tasks: question analysis and question classification.

\subsubsection {Question Analysis}

The goal of question analysis is to gather useful information about the question for the Passage Retrieval module, more concretely: question tokens, token Part-of-speech (POS) tags, headword/compound headword, headword/compound headword synonyms and the headword lexicon target (headword category). For this purpose several  NLP tools are used. A tokenizer is used to identify the various tokens that are in the question. The parse tree of the question is obtained using the Berkeley Parser \cite{petrov07} trained on the QuestionBank \cite{questionbank06}. The POS tags are also obtained by this method. 

As for the question headword, its extraction method consists of traversing the parse tree of the question top-down using a set of predefined rules. 

Regarding compound headword (combination of the question headword with more words that constitute a single unit of meaning), its method of extraction consist of identifying a compound word that includes the headword of the question. The compound headword helps to classify the question into a finer grained category.

For questions without headword which are harder to classify properly a set of lexical and syntactic patterns are used assign to a category to these questions. 
%These same lexical and syntactic patterns have are used in two other different contexts:  to assign a category to questions that have explicit information about the type of answer that is expected -- \emph{“How old is the Empire State Building?”}; to determine the focus of non-factoid questions and questions about acronyms and counts.
The headword category is obtained using a set of hand-written rules to map the headword to a category.

\subsubsection {Question Classification}

Question classification purpose is to assign a predefined semantic category that represents the expected semantic type of the answer. Just.Ask utilizes the Li and Roth \cite{li02} two-layer question taxonomy and to classify questions uses three types of features: lexical, morpho-syntactic and semantic features.

For lexical features \emph{n}-grams and stems are used. Just.Ask uses Unigrams, Bigrams and Trigrams and each word level \emph{n}-gram is used as binary feature, indicating that the question contains the  \emph{n}-gram or not. A stem is the grammatical root of a certain word. Stemming is therefore, a technique that reduces words to their correspondent stems (for example, after stemming playing and played would become play). Just.Ask uses stemming after representing the question using  \emph{n}-grams, to transform each word of a  \emph{n}-gram into its corresponding stem. Stemming is used in combination with the removal of stop words, to reduce the total number of features that have to be considered by the classifier.

Regarding morpho-syntactic features, POS tags and the question headword are used. The process of extraction and use of these features was already described in the previous section. 

Named entities and semantic headword are used as the main semantic features. Just.Ask takes advantage of a Hidden Markov Model based entity recognizer provided by LingPipe, to extract named entities from questions. After the extraction, the recognized named entities can be added as features to the classifier (feature enrichment) or can replace the identified named entities with the correspondent entity category (feature reduction). The semantic headword consists of a improving the headword feature by enriching it semantics. For that purpose Just.Ask utilizes Wordnet \cite{fellbaum98} lexical hierarchy to associate a higher-level semantic concept to a question headword. 

As for the classification itself, it can be modeled according to two types of techniques: hand-built rules or machine learning techniques like Support Vector Machines (SVM), K-Nearest Neighbors (KNN) and Naive Bayes.

\subsection {Passage Retrieval}

The Passage Retrieval module receives an interpreted question from Question Interpretation Module and outputs the relevant passages (passages which are more likely to contain an answer) for that question. Based on the question category, Just.Ask uses different information sources (like Google and Wikipedia) and different query formulation techniques (according to the type of information source being used). Google and Yahoo web search engines are used to answer factoid-type questions, while Wikipedia and DBpedia are utilized for answer non-factoid type questions. After receiving the query results from an information source their metadata, such as the rank (importance of each search result) is locally store for future use.

\subsection {Answer Extraction}

The Answer Extraction module receives the interpreted question and the relevant passages from the Interpretation and Passage Retrieval Modules respectively, and returns the answer to the user question. This module performs two main tasks:  candidate answer extraction and answer selection.

Candidate answer extraction uses different extraction techniques according to the question type taxonomy utilized in the question classification step. Regular expressions are utilized as an extraction method for NUMERIC type questions. A machine learning-based named entity recognizer is used for HUMAN:INDIVIDUAL type questions. There are questions like ``\emph {Which animal is the slowest?}'', in which the answer is not an instance of animal but a type of animal. To extract candidate answers from this type-of questions Just.Ask uses WordNets hyponymy relations, since the candidate answers for these questions are often hyponyms of the question headword. For LOCATION:COUNTRY and LOCATION:CITY type questions it is used a gazetteer (a geographical dictionary) to help extract the candidate answers.

Regarding answer selection, this process is done in three steps. First one is Candidate Answer Normalization, as the name says, consists of reducing to a canonical representation answers that belong to the categories NUMERIC:COUNT and NUMERIC:DATE. The next step is Clustering, which consist of grouping similar answers together according to some similarity criteria (the overlap distance or the Levenshtein distance). Each cluster has a score assigned to it, in which in the most of the cases corresponds to number of members in the cluster. Also each cluster has a answer representative that is defined as the longest answer in the cluster. The final step is Filtering, as the name seems to indicate, it consists in removing undesired answers according to a certain filtering criteria. In this case, if any of the answers present in any of the clusters is contained in the original question, than the whole cluster is discarded. After this filtering process the representative answer of the cluster with the highest score is returned.

\section {Aranea Architecture}

Although some modules names are different (but at the functional level their responsibilities are the same),  the Aranea system possesses the same pipelined architecture as Just.Ask (as showed in Figure 2). Aranea is a system based on two different approaches, one searches answers in structured resources and the other uses data redundancy of unstructured informations sources (like the Web) to do so. The data redundancy approach is the main approach used in Aranea and his most known feature. 

\begin{figure}[h!t!]
\centering
\includegraphics[scale=0.4]{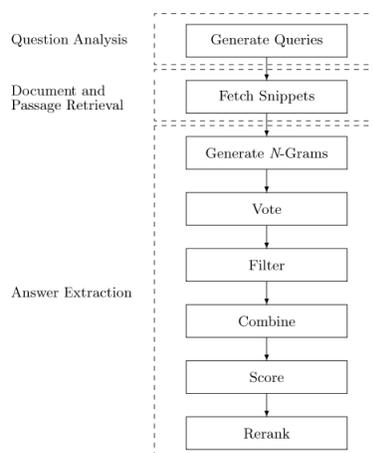}
\caption{Aranea Architecture}
\label{fig:Aranea_Architecture}
\end{figure}

\subsection {Question Interpretation}

Its main purpose is to formulate the queries that will be sent to the Passage Retrieval module.  Besides the baseline query (the question in natural language submitted by the user) Aranea uses two types of queries: exact and inexact queries. 

An exact query is a reformulation of the baseline query that anticipates the specific location of an answer and tries to extract it. Exact queries are generated using pattern matching rules based on question terms and their POS tags (matches are done at the morpho-lexical level).  For instance, the answer to the question ``\emph {Where is The Grand Canyon?}'' will probably appear next to the phrase ``\emph {The Grand Canyon is located in}''. Therefore, the correspondent exact query of that question is \emph ``{The Grand Canyon alto is located in ?y}'', where ?y indicates the location of the anticipated answer.

An inexact query is also a reformulation of the baseline query and it is generated by the same pattern matching rules of exact queries. Thus, each time a pattern is matched at the morpho-lexical level exact and inexact queries are generated. Inexact queries are identical to exact queries without the unbound variable but the query terms are treated as a bag of words. Due to the latter property inexact queries present the benefit of a broader coverage since an exact match of the inexact query is not required. The inexact query for the above example would be \emph ``{The Grand Canyon is located}''.

To extract answers to questions Aranea uses two methods based on the data redundancy approach: exploiting the statistical associations between question and answer terms and extracting answer using reformulation patterns.  The first is accomplished by the use of baseline and inexact queries, and the second is accomplished by the utilization of exact queries.

\subsection{Passage Retrieval}

The goal of this module is to retrieve passages that are likely to contain an answer to the user question. For that purpose, the queries created in the previous module are translated by Aranea into valid Google and Teoma queries, and are sent to these search engines. Aranea operates only on the snippets returned by the search engines, since these small text segments represent relevant passages extracted from the retrieval pages. An amount of 100 snippets are obtained (when possible) for each query, and a score is attributed to these snippets based on type of the query (exact queries are given five times more weight than the other types). 

\subsection {Answer Extraction}

Its main purpose is to give the user the final answer to his question. This module has several tasks that are performed by various sub-modules that will be described in the following sections.

\subsubsection {Generating \emph {n}-grams}

The system uses \emph {n}-grams to serve as initial candidate answers. Aranea generates Unigrams, Bigrams, Trigrams and Tetragrams from the snippets obtained by the Document and Passage Retrieval module. To these \emph {n}-grams are assigned a score that is equal to the score of the query that originated them.

\subsubsection {Voting}

A new score is attributed to each \emph {n}-gram (candidate answer) in which the new score is the sum of the scores of all occurrences of that \emph {n}-gram in the snippet. So n-grams that occur more frequently have higher scores and have higher chances of being correct answers. However, some of top-scoring candidate answers are stop-words which are not decent candidate answers.

\subsubsection {Filtering}
\label{sec:Filtering}

In order to deal with previous mentioned problem, Aranea applies three types of filters: type-neutral, type-specific and closed-class.  This is not a perfect filtering process, since the filtering does not eliminate all of the wrong candidate answers (these will be eliminated in future steps).

The neutral type filter implements two heuristics that are neutral to the question type.  The first one discards candidate answers that begin or end with stopwords. The second one discards candidate answers that contain words that are found in the user question. This heuristic is not applied for questions in which question focus words are likely to appear in the respective answer.

The specific type filter as the name suggests implements several rules that are specific with respect of the question type. For example,  ``\emph {How old?}'', ``\emph {How long?}'', ``\emph {How hot?}'' type questions have a numeric component (in the form of numeric digits or written numbers), therefore answers that do not contain this numeric component are discarded.

The closed-class filter is applied to questions whose possible answers can be enumerated because they are part of a restricted domain. For instance ``\emph {What nationality has Conan O’brien?}'' must be answer with a nationality, so answer candidates that are not known nationalities are discarded. Aranea implements filters for 17 different answer types. Some of these answer types are capitals/population of countries, the meaning of a certain acronym and the symbol of a certain chemical element. Answer types may also include specific information about states of the U.S.A (like area, mottoes, creation date etc) and presidents of the U.S.A (like when the respective presidency term began, the order in the list of U.S.A presidents a certain president is etc). 

\subsubsection {Combining}

As a consequence of the \emph {n}-gram generation shorter frequent candidates tend to be favored. In order to diminish this effect unigrams are used to increase the score of longer answer candidates. More specifically, the score of a candidate answer is incremented by sum of the scores of its unigram components.

Three techniques are used to prevent the increase of score for long answer candidates that contain words that have no importance to the answer. One of them consists of adjusting the score of a candidate based on the frequency of his occurrence (less frequent candidates will have lower scores). The remaining two are the filtering techniques already described and the inclusion of inverse document frequency in the formula that calculates a candidate answer score.

\subsubsection {Scoring}

Answer candidates with less common terms should be preferred over answer candidates with more common terms (since in the Web terms do not have the same occurrence frequency).

Based in this fact, in order to improve the quality of candidate answers, Aranea incorporates the inverse document frequency in the score calculation formula. Statistics collected from the AQUAINT collection are used as ``distribution model'' for the distribution of the terms on the Web. The following formula represents the final formula for calculating the score of an answer candidate.

$$ S_c = S_c \times \frac {1} {|c|} \sum_{w \in c}^{} \log {\bigg (\frac {N} {W_{cnt}}\bigg )}$$

The variable c denotes the set of terms in the candidate answer; N is the total number of documents in retrieved collection and Wcnt is the number documents where word w appears. 

\subsubsection {Reranking}

The functional purpose of this step is similar to the one of the type-specific filter described earlier. A set of heuristics are used to detect likely correct answer for specific question types. The recognized answer forms will be promoted to the top of the ranked list of answers. Before giving the final answers, Aranea counts the number of snippets that contain the answer candidates and discards answer candidates which are not contain by at least two snippets. If all candidate answers are discarded by this process the system returns ``don’t know''.

\section {Open Ephyra}

Opeh Ephyra also shares the same high-level architecture as the previous systems. However, the main modules use many different techniques from the ones previously described in the last two sections to accomplish their purposes (as it is shown in Figure 3). The main characteristic of Opeh Ephyra is the utilization of Patterns Learning techniques to interpret questions and to extract answers. Also the system is easily extended and highly adaptable to other languages. This last property was achieved by performing a separation of language-specific from language-independent code and by defining patterns that are specific for the English language in separate resource files.

\begin{figure}[h!t!]
\centering
\includegraphics[scale=0.5]{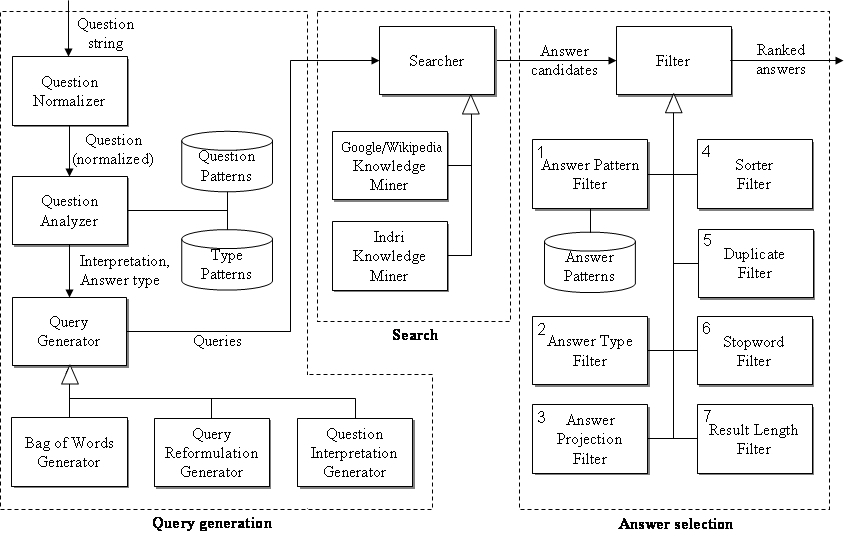}
\caption{Open Ephyra Architecture}
\label{fig:Ephyra}
\end{figure}

\subsection {Question Interpretation}
\label{sec: Question Interpretation}

This module receives the user question as input and outputs queries that will sent to Google and Yahoo search engines. The “Question Normalizer” sub-module generates two question representations: one is used in question analysis and the other is utilized in the generation of queries for the Passage Retrieval phase.  In the first representation all verbs are replaced by their infinity form and all nouns are reduced to the singular form. In the second representation the verb constructions are replaced by the verb construction that is expected to appear in the answer (for example ``When did Shakespeare write Hamlet?" is transformed in ``Shakespeare wrote Hamlet ?''). In both representations unnecessary punctuations marks and words/phrases are eliminated.

The first question representation is utilized by the Question Analysis sub-module to obtain an \emph{interpretation} of a question. This \emph{interpretation} is extracted using a set of manually defined question patterns and allows formulating a query that preserves the question semantics. The extraction patterns are based on the assumption that the \emph{interpretation} of a question can be reduce to three components -  A question ask for a \emph{property} of a \emph{target} in a specific \emph{context}. After the extraction of these components the Question Interpretation Generator then transforms the \emph{interpretation} into a query string.

The Question Analysis sub-module is also responsible for the question classification. Ephyra uses a set of 154 answer types arranged in a hierarchy with 44 top-level categories, which can be consulted in \cite{PaperEphyra:2007}. Like Just.Ask, Ephyra uses lexical, syntactic and semantic features for question classification.  Unigrams and bigrams are the only lexical features utilized. As syntactic features are used: the focus adjective or adverb of the question (only applicable for ``\emph{How}'' questions), the main verb of the question, the question word and another feature that indicates whether the question word serves as determiner of the focus word. The semantic type of the question focus word (only applicable to``\emph{What}'' or ``\emph{Which}'' questions), is the only semantic features utilized. The focus word is identified with a set of manual syntactic patterns. The semantic type of focus word is determined using the Wordnet. 
As for the classification itself, it Ephyra possesses three different type classifiers: rule-based, model-based and hybrid (combines the outputs of the previous classifiers using their associated confidence scores).

Using the second question representation, four types of queries are generated. The Bag of Words sub-module creates queries  based on the most important content words of the question - Keyword queries. The Question Reformulator sub-module generates queries that correspond to possible formulations of what the answer sentence might be (like Aranea does with inexact queries) - Reformulation queries. Although not depicted in Figure 3,  Predicate queries are formed from the predicate verb and its arguments by the Predicate submodule and Term queries which consist of the question terms (single tokens or expressions) expanded with semantically related concepts are created by the Term sub-module. 

\subsection {Passage Retrieval}

For factoid and list type questions Ephyra can use Google or Yahoo as sources of information to retrieve relevant passages. All the previous described query types are sent either Google or Yahoo to accomplish that goal.
% For other type questions besides Google Wikipedia is also used as an information source (the next section explains the 
% motivation for the use of Wikipedia in more detail).

%Indri is used to retrieve passages from the AQUAINT2 corpora, and uses all the previous described query types except %the ``formulations” query type.   

\subsection {Answer Extraction}

According to the different types of questions, different strategies for answer extraction and selection are utilized.

\subsubsection {Factoid, Definition and List Questions}
\label{sec:Answer Extraction Approaches}

Two approaches can be used for extracting candidate answers from these question types: Answer Type and Answer Pattern. Each one of these approaches is implemented by the corresponding filter presented in Figure \ref{fig:Ephyra}.
The Answer Type Filter applies a NE tagger to extract candidate answers of all the possible answer types. The Standford NE recognizer \cite{StandfordParser:2005} extracts answers of the types \emph{person}, \emph{organization} and \emph{location}, rule- and list-based taggers extract answers of the remaining types. The extracted NEs are tokenized and stemmed to identify similar NEs. For each cluster of similar entities, one representative is chosen and it is assigned to it a score equal to the number of entities in the cluster. As we can see this approach is much similar to the one approach used in Just.Ask for the same purpose.

The Answer Pattern Filter uses a set of patterns to extract answers from text snippets that possess both the target object and the desired property of the target. These patterns can be hand-built or can be automatically learned using question-answer pairs as training data.
Each question is interpreted and a query string with the interpretation of the question and its answer is formed. The query string allows to fetch text snippets that contain answer patterns for a specific property. An answer pattern contains the target, the property, an arbitrary string in between these objects and one token preceding or following the property (to indicate where it starts or ends). For instance, for the property profession the following answer pattern could be found: \textless Target\textgreater works as \textless Property\textgreater. The score of the candidate answers is calculated using a statistical approach that cumulate the confidence measures of the answer patterns used to extract them (a detailed description can be found in \cite{PaperEphyra:2006}).

% Since this calculation method works better for a large and redundant knowledge source, answers are first extracted from % the Web (using Google) and then they are projected into the AQUAINT2 corpus. Both answer type and pattern matching % extractions techniques are applied to the corpus to determine documents that support the answers from the Web.   %Documents found with the pattern matching approach are preferred to those found with the answer type approach. 
% If  neither approaches detect the Web answer in the corpus, the hit with bigger relevance returned by Indri is used as %the supporting document.

%Although not present in Figure 3, a third answer candidate extractor technique can be utilized. This extractor consists in a semantic parser that generates a semantic representation of the question and extracts answer candidates from phrases in the corpus. %Similarity measures between the semantic structures of questions and the relevant passages of the corpus, are utilized as criteria for extracting answer candidates. 

Answer selection is performed by applying consecutively a set of filters to the candidate answers to improve their quality (by eliminating worse candidates answers according to a certain criteria). This filtering process works as follows: First, the Sorter Filter arranges the extracted candidate answers according to their score. After this, the Duplicate Filter compares the answers pairwise. When it detects two similar answers, it drops the lower ranked answer and adds its score to the score of the higher ranked one. Answers are considered similar, if they contain content words that have the same stem. The Stopword Filter discards answers that are not in the correct format. The Result Length Filter ensures that the list of answers does not exceed the maximum number of 7000 non-whitespace characters. Also, all answers having scores below a defined threshold are discarded. If the final list of answers is empty, ”NIL” is returned to indicate that the system could not find an answer.

\subsubsection {Other Questions}

Open Ephyra can also deal with questions who are not of the factoid and list type. However, due to the motives that are going to be discussed in Section 7 the processing pipeline for this type of questions is not in the scope of this work and thus will not be described in this section. The processing pipelining for these type of questions can be consulted in \cite{PaperEphyra:2007}.     

\section {Preliminary Evaluation}

\subsection {Just.Ask Characteristics}

Just.Ask can only receive as input factoid, definition and explicative questions. As output the system returns the three best answers to the question at hand. For each answer it also returns part of the snippet where that answer was found and the web-address of that snippet.

\subsubsection {Corpus}

Just.Ask was evaluated using a corpus with 200 questions and the respective correct answers for those questions, the gold-QA corpus. This corpus is an extended version of the Multisix corpus that suffered some manual modifications in order to: assure that all questions are valid, there is a good variety of questions and classify all questions according to the Li and Roth question type taxonomy. Table 1 shows the number of questions in the corpus according to the question word.Questions in this corpus have the following format:

Question 1. Q:What is the capital of Somalia - \{Mogadishu Somalia\} \{Mogadishu\} - LOCATION\_CITY.

So questions are composed of: a question Id, the question itself, a set of possible answers (each answer is between brackets) and the question classification.

\begin{table}
\begin{center}
\begin{tabular}{|c| c c c c c c c c |c|}
\hline
Question word & Who & When & Where & What & Which & How & Name & Other \\ 
\hline 
\# Questions & 36 & 25 & 22 & 50 & 8 & 29 & 8 & 22 \\ 
\hline 
\end{tabular}
\newline 
\caption{Number of questions in our test corpus per question word.} 
\end {center}
\end{table}

Just.Ask uses search engines to search the web for answers to questions, in order to protect the validity of the answers, these are searched in a local corpus – the Web Corpus. This corpus contains passages that were collect from queries sent to Google for each question present in the gold-QA corpus. The possible answers for each question in gold-QA were updated in order to match the answer possibilities conveyed by the Web Corpus.
The format of passages is too extensive to put a concrete example here, however they contain: the question Id, the question, the URL of passage, the title that is showed in that URL, a short description of the content of that URL, the passage rank by the system, and a flag that indicates if the passage were retrieved from Wikipedia of DBpedia. 

\subsubsection {Baseline Results}

Using this setting an extrinsically evaluation was performed. The best results were achieved using only 32 passages from the ones that were collected from Google. Table 2 presents the overall system results according to several measures.

\begin{table}[h!t]
\begin {center}
\begin{tabular}{|ccccc|}
\hline
\multicolumn{2}{|c||}{\# Questions} &  Correct & Wrong & No Answer   \\
\hline
\multicolumn{2}{|c||}{200} & 88 & 87 & 25 \\ \hline \hline Accuracy &  Precision & Recall & MRR & Precision@1\\ \hline 44\%  & 50.3\% & 87.5\% & 0.37 & 35.4\%\\ \hline \end{tabular} \caption{Just.Ask best results.} \label{table:overallresults} \end{center} \end{table} 

\begin{table}[h!t]
\begin {center}
\begin{tabular}{|c|cccc|}
\hline
Question word & Correct & Incorrect & No answer & Accuracy\\
\hline
Who & 23 & 13 & 0 & 63.89\%\\
When & 15 & 10 & 0 & 60.00\%\\
Where & 10 & 12 & 0 & 45.50\%\\
What & 19 & 16 & 15 & 38.00\%\\
Which & 3 & 4 & 1 & 37.50\%\\
How & 8 & 18 & 3 & 27.59\%\\
Name & 3 & 2 & 3 & 37.50\%\\
Other & 7 & 12 & 3 & 31.82\%\\
\hline
Total & 88 & 87 & 25 & 44.00\% \\
\hline
\end{tabular}
\caption{Just.Ask results according to the different question words.}
\label{table:resultsqwords}
\end {center}
\end{table}

Table 3 presents the systems results according to the different question words. As can be seen by the last table, Just.Ask has a good performance in questions that typically involves the name or definition of a person (\emph{What} and \emph{Who} questions). \emph{How} questions seem to be challenging for the system since they possess the worst results.

\subsection {Aranea Characteristics}

As stated in \cite{Jimmy:2007}, Aranea can only deal with factoid type questions. As output the systems returns N possible answers according to the criteria discussed in Section 4.3.6.

\subsubsection {Corpus}

Aranea was evaluated by running the system in the TREC-9, TREC-10, TREC-11 and TREC-12 questions data sets, respectively. These questions sets were manually modified in order to contain only factoid questions. The question format\footnote{\url {http://trec.nist.gov/data/qa/T9_QAdata/qa_questions_201-893}} is very simple and only contains the question id, the type of the question and the question itself. Evaluations of TREC-9 and TREC-10 questions sets use the TIPSTER\footnote{\url{http://www.nist.gov/tac/data/data_desc.html}} collection as corpus. The documents of that collection were tagged using SGML (Standard Generalized Markup Language). Evaluations of TREC-11 and TREC-12 use the AQUAINT\footnotemark[\value{footnote}] collection as corpora. Its documents were also tagged using SGML. Besides the local corpus, the web is also utilized as corpus in all evaluations. But since Aranea implements a caching mechanism that allows the reuse of previously fetched Web pages, all runs can be replicated (without the issue of obtaining different results in each run).

\subsubsection {Baseline Results}

The Aranea default configuration in which he uses all query types and retrieves at most 100 snippets per query, was evaluated in setting previously described. Three evaluation measures are used: Mean-Reciprocal-Rank (MRR), the fraction of questions with correct answers at rank one (C@1) and the fraction of questions for which a correct answer was found in top five returned answers (C@5). The evaluation results can be consulted in Table 4 . The use of both Google and Teoma engine searches outperforms the use of either one of the engines individually.

\begin{table}
\begin{center}
\begin{tabular}{|c| c| c| c|}
\hline
& MRR & C@1 & C@5 \\ 
\hline 
Both & 0.537 & 0.477 & 0.630 \\ 
\hline 
Teoma & 0.508 & 0.454 & 0.591 \\
\hline
Google & 0.495 & 0.441 & 0.581 \\
\hline
\end{tabular}
\newline 
\caption{Average of Aranea's performance in all TREC data sets} 
\end {center}
\end{table}

\subsection {Open Ephyra Characteristics} 

Like Just.Ask Open Ephyra can receive as input factoid, list and definition question types. As output the systems returns N possible answers according to the criteria discussed in Section 5.3.1.

\subsubsection {Corpus}

Open Ephyra was evaluated by running the system in the TREC-2007 question data set.  Although the question data set file is written in XML the questions have the format as the questions that Aranea uses. This evaluation uses both the AQUAINT2 and BLOG06 collection\footnote{\url{http://www.nist.gov/tac/data/data_desc.html}} as local corpus. AQUAINT2 documents were tagged using XML and BLOG06 documents possess the raw HTML content from blogs all over the Web wrapped between a \textless DOC \textgreater...\textless /DOC \textgreater pair. Besides these collections the web is also utilized as corpus in the evaluation. 
For answering factoids and list questions answers are extracted from the Web and then are projected onto the local corpus. 

\subsubsection {Baseline Results} 

Two runs of the system were executed in the described evaluation setting with slightly modifications. \emph{Run1} used only the AQUAINT2 corpus, while \emph{Run2} combined the AQUAINT2 and BLOG06 corpora and treated them as unique source of information. Both \emph{Run1} and \emph{Run2} use the statistical approach for selecting answers from answer candidates. Regarding the evaluation measures utilized, the F1 measure can be interpreted as an equally weighted average of precision and recall. The F3 measure in its turn weights recall higher than precision. The results of these evaluations are described in Table 5. 
Analyzing the results the use of both AQUAINT2 and BLOG06 corpus leads to the best results for list and factoids questions. Also recall is higher when only using AQUAINT2 as corpus.

\begin{table}
\begin{center}
\begin{tabular}{|c| c| c|}
\hline
& \emph{Run1} & \emph{Run1} \\ 
\hline 
Corpora & AQUAINT2 & AQUAINT2, BLOG06 \\ 
\hline 
Unsupported (U) & 28 & 23 \\
Inexact (X) & 18 & 23 \\
Locally Correct (L) & 1 & 1 \\
Factoid Accuracy & 0.206 & 0.208 \\
\hline
F1 measure & 0.140 & 0.144 \\
\hline
F3 measure & 0.189 & 0.156 \\
\hline 
Average per-series score & 0.181 & 0.172 \\
\hline
\end{tabular}
\newline 
\caption{Results of Open Ephyra evaluation} 
\end {center}
\end{table}

\section {A Common Evaluation Setting}
\label{sec: Common Evaluation Setting}

The previous section discussed for each QA system in study, the most important characteristics to take into account in their evaluation – type of processed questions, the corpus utilized, evaluation measures used. Based on these characteristics a common evaluation setting can be devised to evaluate all systems in equal terms.

The first aspect that has to be taken in consideration for that purpose, is that the systems can only be evaluated in equal terms for question types that all systems process. Aranea is the most restrictive system since it only processes factoid questions, so all systems can only be evaluated in the processing of this question type. Based on this assessment, gold-QA seems to be the best option as question corpora for the common evaluation setting. This decision is motivated by the fact that gold-QA is composed in its majority by factoid questions and has a smaller set of other question types (that have to be manually removed later) compared to the question corpus of Aranea and Open-Ephyra.

Since gold-QA is going to be utilized as question corpus, choosing Web Corpus as the corpus for answers would probably be the best option due to the fact that Web Corpus is highly adapted to the contents of gold-QA, possessing an average of 3.89 answers per question.  Unfortunately this is not a viable option, since it would not allow to do proper evaluation of the three systems at hand. For understanding this assertion, it must be remembered that Web corpus was built with passages obtained from Just.Ask keyword queries sent to Google. So in order to study the contribution of the several query types used by Aranea and Open-Ephyra for the QA performance, the passages obtained by these query types must be utilized. Passages contained in Web Corpus cannot be used for that purpose since they do not represent faithfully the passages retrieved by those query types. Due to this fact, the Web will be used as the corpus for answers. This decision has some drawbacks, namely the non reproducibility of obtained results (in other words, the execution of a test in two different periods of time is likely to produce different results). To reduce this ``inaccuracy'' in results to a minimum, each test that will be conducted is going to be run 5 times and the final results of that test are the average of the results obtained in all runs.   

Regarding the evaluation measures, part of the Just.Ask evaluation measures will be used for that purpose. These measures are the following:

\textbf{Accuracy}, defined as the number of correctly judged items, divided by the total number of items in the test corpus i.e,

$$ Accuracy =\frac {\# Correctly\,judged\,items} {\#Items\,in\,the\,test\,corpus}$$

\textbf{Precision}, defined as the number of relevant items retrieved, divided by the total number of items retrieved, i.e,

$$ Precision =\frac {\# Relevant\,items\,retrieved} {\#Items\,retrieved}$$

\textbf{Recall}, defined as the number of relevant items retrieved, divided by the total number of relevant items in the test corpus, i.e,

$$ Accuracy =\frac {\# Relevant\,terms\,retrieved} {\#Relevant\,items\,in\,the\,test\,corpus}$$ 

\textbf{Mean Reciprocal Rank} (MRR), used to evaluate systems that return ranked lists of items to a query. The reciprocal rank of an individual query is defined to be the multiplicative inverse of the rank of the first relevant item, or zero if no relevant items are returned. Thus, the mean reciprocal rank is the average of the reciprocal rank of every query in a test corpus. More formally, for a test collection of N queries, the MRR is given by:

$$ MRR =\frac {1} {N}  \sum_{i = 1}^{N} \frac {1} {rank_i}$$

\textbf{Number of Positive Passages}, defined as the number of passages that contain the answer to a certain question.

Summarizing the contents of this section, the evaluation setting where all three system will be compared has gold-QA as the question corpus, the Web as answer corpus and the previously described evaluation measures are used for evaluate the performance of the systems. 

The next section will explain the rationale beyond the decision of evaluation Aranea apart from Just.Ask and Open-Ephyra, and why the scope of the evaluation of Just.Ask and Open-Ephyra will also include the evaluation of definition questions (this will be the setting used in the evaluations perfomed in Sections\ref{sec: Just.Ask and Open-Ephyra Extrinsic Evaluation} and  \ref{sec: Just.Ask and Open-Ephyra Intrinsic Evaluation}).

\section {Adaptation to the Common Evaluation Setting}
\label{sec: Adaptation to Evaluation Setting}

The necessary changes in all systems were made in order for the systems to be evaluated in the defined evaluation setting, more specifically, the systems had to be adapted to process gold-QA corpus and the evaluation measures had to be implemented in each one of the systems. After these adaptations were made, a few preliminary tests that were conducted to assure that the systems could be evaluated in the defined setting showed some problems with Aranea and with Open-Ephyra.

Regarding Aranea, although the system seemed completely operational it only gave answers to a few specific question types (described in Section \ref{sec: Aranea Evaluation}). Besides this fact, questions that according to system documentation should produce a correct answer, produced no answer at all. A depth analysis of the system code was performed to understand the source of this problem that causes a severe deterioration in system performance, however, with no success. The system authors were also consulted to provide advice in solving this problem. Unfortunately, no piece of advice was provided by the authors since they stopped doing research with Aranea five years ago. Due to the incapability of solving this issue, Aranea produces much worse overall results when compared to Just.Ask and Open-Ephyra. For that reason, the evaluation of Aranea will be described in a different section from the evaluation of the remaining systems. Since Aranea will be evaluated in a more ``restricted setting'' the evaluation of Just.Ask and Open-Ephyra no longer needs to be restricted to the evaluation of factoid questions. So the evaluation of these two systems will have has scope factoid and definition questions.
%and the comparison with Just.Ask and Open-Ephyra will have its own section.  

As for Open-Ephyra, although the system is described as capable of using Google as information source, in practice that is not possible. Every time query is sent to Google a error occurs related with the type of arguments that are passed to the Google API. This fact seems to indicate that the error occurs due to an update in Google's API to which the necessary modifications were not made in Open-Ephyra to work with the new API.  Nevertheless this incapability of using Google it is not restrictive to the evaluation of Open-Ephyra and Just.Ask, since they both can be evaluated in equal terms if Yahoo is used as information source. 

\section{Aranea Evaluation}
\label{sec: Aranea Evaluation}

The best results of Aranea were achieved using the system default setting - Google is used as information source and the system tries to obtain a 100 passages per query. Table \ref{table:AraneaResults} presents the system overall results. In a total of 200 questions, Aranea only attempted to answer 7 questions. From these, 4 were correct and 3 were wrong. Being so, the system attained 43\% of precision in those 7 questions and 2\% of accuracy regarding all the questions in the question corpus.

\begin{table}
\begin{center}
\begin{tabular}{|c| c|}
\hline
Measures & Results \\  
\hline 
Correct Questions & 4 \\
\hline
Wrong Questions & 3 \\
\hline
No Answer & 193 \\
\hline
Accuracy & 2\% \\
\hline
Precision & 43\% \\
\hline
Recall & 3.5\% \\
\hline 
\end{tabular}
\newline
\caption{Aranea extrinsic evaluation}
\label{table:AraneaResults} 
\end {center}
\end{table}

The 7 questions that Aranea attempted to answer are:
\begin{itemize}
\item ``What is the capital of Somalia?";
\item ``What is the capital of Alaska?";
\item ``What is the capital of Chechnya?";
\item ``Who was the third U.S. President?";
\item ``When did Alaska become a state?";
\item ``What does the acronym "PERS" stand for?";
\item ``What is the capital of Madagascar?";
\end{itemize}

As it can be seen these questions ask information regarding: a president of the U.S.A, capitals of several countries, creation date of an American state and the meaning of a certain acronym. Aranea uses the closed-class filter(described in sub-section \ref{sec:Filtering}) to filter answer candidates for these question types. In order to see if this filter can be used in Just.Ast and Open-Ephyra to improve the results to these question types, these 7 questions were submitted to the systems. In this experiment, Just.Ask and Open-Ephyra use the same setting as Aranea, except they used Yahoo as information source (due to the problem described in the previous section). Table \ref{table:AraneaComparison} shows the comparison of the performance of the three systems for those questions.

\begin{table}
\begin{center}
\begin{tabular}{|c| c| c| c|}
\hline
& Aranea & Just.Ask & Open-Ephyra \\  
\hline 
Correct Questions & 4 & 7 & 7 \\
\hline
Wrong Questions & 3 & 0 & 0 \\
\hline
\end{tabular}
\newline
\caption{Comparison between Aranea, Just.Ask and Open-Ephyra}
\label{table:AraneaComparison} 
\end {center}
\end{table}

Just.Ask and Open-Ephyra clearly outperform Aranea for these questions by answering them all correctly. Based on these results it can be concluded that techniques used by Just.Ask or Open-Ephyra answer to these types of questions could be used in Aranea to improve the system performance. However it is hard to specify which techniques from these two systems could be useful for Aranea, since both systems do not have a specific approach towards these types of questions.

As a final remark of this section, it is important to say that no intrinsic evaluation and comparison of Aranea's modules with the respective modules of the remaining systems is performed, due to problems in obtaining important data to do those tasks  (namely, passages retrieved and candidate answers selected by the system). 

\section{Just.Ask and Open-Ephyra Extrinsic Evaluation}
\label{sec: Just.Ask and Open-Ephyra Extrinsic Evaluation}

This section evaluates and compares the overall results of Open-Ephyra and Just.Ask according to different parametrizations in the number of retrieved passages and the utilization/non-utilization of features of the systems. All these evaluations and comparisons will be performed using the setting defined in Section \ref{sec: Common Evaluation Setting} and will use Yahoo as information source (due to the reasons mentioned in Section \ref{sec: Adaptation to Evaluation Setting}). 

\subsection{Best Performance}

In their default setting\footnote{We understand by default setting, the set of parametrizations that a system possesses when it is obtained from the respective authors.} Open-Ephyra retrieves up to 800 passages per query, while Just.Ask needs a specific parametrization of the maximum number of passages that a query can retrieve.  Previous evaluations of Just.Ask performed in \cite{Just.Ask:2009} indicate that Just.Ask performs best when using 64 as the maximum number of passages to retrieve. Experiments were conducted in order to verify if this assessment is valid for the present evaluation setting. Since the assessment is valid in the current evaluation setting in this and following sub-sections Just.Ask is going to be evaluated using 64 passages as the maximum number of passages to retrieve. 

Several experiments were conducted to perceive if the default setting of Open-Ephyra (the 800 passages per query) attains the system best results. When increasing the number of passages retrieved, no improvement of the results were verified. The decreasing of the number of passages retrieved  led to some interesting results. When using more than 100 passages per query the default settings results remain unchanged, while when using less that 50 passages the results suffered a major degradation. Additional experiments led to the conclusion that 64 passages is the lowest number of passages that can be utilized to achieve the system best results. Since both Just.Ask and Open-Ephyra attain their best results using 64 passages, they can be evaluated and compared under the exact same conditions. The results of the evaluation are depicted in Table \ref{table:Just-Ephyra-Extrinsic-Results}.

\begin{table}
\begin{center}
\begin{tabular}{|c| c| c|}
\hline
\multicolumn{3} {|c|} {64 Passages}\\
\hline
Measures & Just.Ask & Open-Ephyra \\  
\hline 
Correct Questions & 73 & 79 \\
\hline
Wrong Questions & 103 & 107 \\
\hline
No Answer & 24 & 14 \\
\hline
Accuracy & 36.5\% & 39.5\% \\
\hline
Precision & 41.5\% & 42.4\% \\
\hline
Recall & 88\% & 93\% \\
\hline 
\end{tabular}
\newline
\caption{Just.Ask and Open-Ephyra extrinsic evaluations}
\label{table:Just-Ephyra-Extrinsic-Results} 
\end {center}
\end{table}

It can be seen by the table that Open-Ephyra has slightly better performance than Just.Ask. Only in the number of wrong questions answered Just.Ask attains the best performance since Open-Ephyra have 4 more additional wrong questions relatively to number of wrong questions answered by Just.Ask. We can conclude that in overall Open-Ephyra is the best QA system from the analyzed systems (since it is able to answer more questions correctly which is the ``main'' goal of a QA system).

\subsection{Overall Impact of Open-Ephyra Answer Extraction Techniques}
\label{sec: Overall Impact of Open-Ephyra Answer Extraction Techniques}

As it was described in sub-section \ref{sec:Answer Extraction Approaches} Open-Ephyra can use two approaches - Answer Type and Answer Pattern - to extract from obtained passages candidate answers, from which the final answer is going to be selected. To understand the impact of each approach to the performance of the system, the system was evaluated using only one approach to extract answers. Table \ref{table:Open-Ephyra-Extrinsic-Results} presents the results of this evaluation.

\begin{table}
\begin{center}
\begin{tabular}{|c| c| c|}
\hline
\multicolumn{3} {|c|} {Open-Ephyra - 64 passages}\\
\hline
Measures & Answer Type & Pattern Type \\  
\hline 
Correct Questions & 73 & 50 \\
\hline
Wrong Questions & 104 & 64 \\
\hline
No Answer & 23 & 86 \\
\hline
Accuracy & 36.5\% & 25\% \\
\hline
Precision & 41\% & 44\% \\
\hline
Recall & 88.5\% & 57\% \\
\hline 
\end{tabular}
\newline
\caption{Open-Ephyra extrinsic evaluation using each one of the answer extraction approaches}
\label{table:Open-Ephyra-Extrinsic-Results} 
\end {center}
\end{table}

Analyzing the table it becomes clear that the Answer Type approach has a major impact in the performance of Open-Ephyra. When solely used this approach attains slightly worse results when compared to the results of the system depicted in Table \ref{table:Just-Ephyra-Extrinsic-Results}, where both approaches are used to obtain those results. More specifically less 6 correct answers and less 3 wrong answers which consecutively results in lower accuracy, precision and recall.
Another interesting fact is that the use of the Answer Type approach produces almost identical results as the Just.Ask evaluation (Table \ref{table:Just-Ephyra-Extrinsic-Results}). This is probably due to the high similarity between the approaches used in both systems to extract candidate answers. So since the use of both Answer Type and Pattern Type approaches leads to a better performance of Open-Ephyra over Just.Ask, the following hypothesis can be formulated:

 \emph{Hypothesis 1 - Just.Ask performance can be improved by the utilization of the Pattern Type approach.}

\subsection{Overall Impact of Open-Ephyra Query-Types}

Section \ref{sec: Question Interpretation} describes the several types of queries that Open-Ephyra uses. Since both systems only have have one query type in common - Keyword Queries - it is interesting to see from an overall perspective what is the impact of only using Keyword Queries in Open-Ephyra results, and if these results are similar to the one obtained by Just.Ask. Table \ref{table:Ephyra-Queries} shows the results of the evaluation of Open Ephyra using only Keyword Queries and the results of Just.Ask from Table \ref{table:Just-Ephyra-Extrinsic-Results}. 

\begin{table}
\begin{center}
\begin{tabular}{|c| c| c|}
\hline
\multicolumn{3} {|c|} {64 Passages}\\
\hline
Measures & Just.Ask & Open-Ephyra - Keyword Queries \\  
\hline 
Correct Questions & 73 & 72 \\
\hline
Wrong Questions & 103 & 113 \\
\hline
No Answer & 24 & 15 \\
\hline
Accuracy & 36.5\% & 36\% \\
\hline
Precision & 41.5\% & 39\% \\
\hline
Recall & 88\% & 92.5\% \\
\hline 
\end{tabular}
\newline
\caption{Just.Ask and Open-Ephyra extrinsic evaluation using the only query type in common - Keyword Queries}
\label{table:Ephyra-Queries} 
\end {center}
\end{table}

As it was expected, the solely usage of Keyword Queries led to slightly worse results compared to the ones where all query types were used (Table \ref{table:Just-Ephyra-Extrinsic-Results}).  More specifically less 7 correct answers and more 6 wrong answers which consecutively results in lower accuracy, precision and recall. When comparing both columns of the table we can see that the performance of both systems is similar, although Just.Ask attains in overall better results. More specifically Just.Ask has a higher accuracy and precision (due to more correct questions and less wrong questions) however, Open-Ephyra has a higher recall (due to the high number of wrong questions). Since by using several types of queries Open-Ephyra obtains better results than Just.Ask we can formulate the following hypothesis:

\emph{Hypothesis 2 - Just.Ask performance can be improved by the utilization of one or more of the following query types used by Open-Ephyra: Term Queries, Reformulation Queries, Predicate Queries and Interpretation Queries.}

\section {Just.Ask and Open-Ephyra Intrinsic Evaluation}
\label{sec: Just.Ask and Open-Ephyra Intrinsic Evaluation}

This section performs an evaluation of the 3 processing steps of the QA pipeline - Question Interpretation, Passage Retrieval and Answer Extraction - of Just.Ask and Open-Ephyra. The detailed analysis of these steps will allow to validate the hypotheses formulated in the previous section. All evaluations performed in this section use the setting defined in Section \ref{sec: Common Evaluation Setting}, Yahoo as information source and 64 as the number of maximum passages to be retrieved.

\subsection {Question Interpretation}

In this step the only thing that could be evaluated is the Question Classification task. However this evaluation is not going to be performed due to the following reason: Just.Ask and Open-Ephyra use different taxonomies for classifying questions. This fact do not allow the evaluation of both systems under the same conditions, so improvement strategies discovered from this evaluation cannot be applied to the systems. Detailed information about Just.Ask Question Classification performance can be found in \cite{PaperClassificacao}, while a simple analysis of Open-Ephyra Question Classification performance can be found in \cite{PaperEphyra:2007}.

\subsection {Passage Retrieval}

The analysis of this step is performed in order to have a specific of idea of: the number of relevant passages retrieved by each system, the ranking\footnote{The ranking of a passage is a measure of the ``quality'' of the passage, since it indicates the reliability of the source from which the passage was extracted. So a high rank indicates high source reliability.}) of the passages retrieved and what can be done to improve these results.

Table \ref{table:Passage Retrieval} presents for each query performed by each system method the number of questions for each there is at least one positive passage (\#Question$_{1+PosPassage}$), the mean reciprocal rank of the first positive passage for all questions (MRR$_{ALLQ}$) and the mean reciprocal rank of the first positive passage only for the questions for which at least one positive passage exists (MRR$_{QPOSPASSAGES}$). 

\begin{table} [h!t]
\begin{center}
\begin{tabular}{| c | c | c| c| } 
\hline 
\multicolumn{4}{|c|}{64 Passages} \\ 
\hline 
QA System & \#Question$_{1+PosPassage}$ & MRR$_{ALLQ}$ & MRR$_{QPOSPASSAGES}$ \\ \hline 
Just.Ask & 157 (78.5\%) & 0.36 & 0.46 \\ \hline
Open-Ephyra & 171 (85.5\%) & 0.56 & 0.66 \\ \hline
\end{tabular}
\newline 
\caption{Just.Ask and Open-Ephyra passage retrieval results} 
\label{table:Passage Retrieval} 
\end {center}
\end{table}

Through inspection of the table it is clearly that Open-Ephyra outperforms Just.Ask in this step. Open-Ephyra obtains more 7\% of positive passages and both MRR$_{ALLQ}$ and MRR$_{QPOSPASSAGES}$ values indicate that Open-Ephyra obtains passages with an higher ranks relatively to Just.Ask's passages. The difference in the obtained results seems to be related with the several query types used by Open-Ephyra. 

In order to determine the veracity of the last statement another evaluation as performed where Just.Ask parametrization remains the same as above, and Open-Ephyra only uses Keyword Queries. Table \ref{table: Passage R Keyword Query} depicts the results of this evaluation.

\begin{table} [h!t]
\begin{center}
\begin{tabular}{| c | c | c| c| } 
\hline 
\multicolumn{4}{|c|}{64 Passages} \\ 
\hline 
QA System & \#Question$_{1+PosPassage}$ & MRR$_{ALLQ}$ & MRR$_{QPOSPASSAGES}$ \\ \hline 
Just.Ask & 157 (78.5\%) & 0.36 & 0.46 \\ \hline
Open-Ephyra - Keyword Query & 157 (78.5\%) & 0.33 & 0.42 \\ \hline
\end{tabular}
\newline 
\caption{Just.Ask and Open-Ephyra (using only Keyword Queries) passage retrieval results} 
\label{table: Passage R Keyword Query} 
\end {center}
\end{table}

We can see that in this parametrization both systems obtain the same number of positive passages and Just.Ask obtains passages with slightly higher ranks relatively to Open-Ephyra's passages. These results confirm that the results in Table \ref{table:Passage Retrieval}, are indeed obtained obtained due to the several query types used by Open-Ephyra. This fact allows us to state that \emph{Hypothesis 2} is valid.

Since we established that Just.Ask can be improved by the the use of Open-Ephyra's query types, it would be useful to understand which query types have more impact on the performance. Table \ref{table: Query Types} help us in this endeavor, by showing for the 171 positive passages retrieved in Table \ref{table:Passage Retrieval} what were the query types that originated them. The query types are organized by a decreasing order of their impact on the performance. 

\begin{table}
\begin{center}
\begin{tabular}{|c| c|}
\hline
\multicolumn{2} {|c|} {171 Positive Passages}\\
\hline
Query Types & Positive Passages Obtained \\  
\hline 
Reformulation Queries & 60 \\
\hline
Keyword Queries & 43 \\
\hline
Interpretation Queries & 40 \\
\hline
Term Queries & 28 \\
\hline
Predicate Queries & 0 \\
\hline
\end{tabular}
\newline
\caption{Query types of the positive passages retrieved in Table \ref{table:Passage Retrieval}}
\label{table: Query Types} 
\end {center}
\end{table}
 
As we can see Reformulation, Interpretation and Term Queries are important for Open-Ephyra passage retrieval performance, so the implementation of these query types in Just.Ask should lead to similar results to the ones obtained by Open-Ephyra. With this information \emph{Hypothesis 2} is transformed in the following conclusion: \emph{Conclusion 1 -Just.Ask performance can be improved by the utilization of Reformulation, Interpretation and Term queries used by Open-Ephyra.}

\subsection {Answer Extraction}

The analysis of this step is performed in order to have a specific of idea of the performance of the main two stages in this process - Candidate Answer Extraction(CAE) ans Answer Selection(AS) - and what can be done to improve their results.
 
Table \ref{table: Just.Ask answer extraction} and \ref{table: Open-Ephyra answer extraction results} present for Just.Ask and Open-Ephyra respectively, the number of questions for which a certain amount of answers was extracted. For those questions, the tables show the amount in which the CAE stage was successful, and the amount in which the AS stage was successful. We consider the CAE stage as being successful when it extracts at least one correct answer. Similarly, the AS stage is successful when one of the extracted correct answers is selected as final answer.

\begin{table} [h!t]
\begin{center}
\begin{tabular}{| c | c | c| c| } 
\hline 
\multicolumn{4}{|c|}{Just.Ask} \\ 
\hline 
Extracted Answers & \#Question & CAE Success & AS Success\\ 
\hline 
0 & 25 & - & - \\ 
\hline
1 to 10 & 22 & 1 & 1 \\ 
\hline
11 to 20 & 15 & 5 & 4 \\ 
\hline
21 to 40 & 58 & 37 & 27 \\ 
\hline
41 to 60 & 21 & 15 & 14 \\ 
\hline
61 to 100 & 36 & 36 & 30 \\ 
\hline
100 to 150 & 16 & 13 & 11 \\ 
\hline
151 to 230 & 7 & 7 & 6 \\ 
\hline
All & 200 & 114 & 93 \\
\hline
\end{tabular}
\newline 
\caption{Just.Ask answer extraction results} 
\label{table: Just.Ask answer extraction} 
\end {center}
\end{table}

\begin{table} [h!t]
\begin{center}
\begin{tabular}{| c | c | c| c| } 
\hline 
\multicolumn{4}{|c|}{Open-Ephyra} \\ 
\hline 
Extracted Answers & \#Question & CAE Success & AS Success\\ 
\hline 
0 & 15 & - & - \\ 
\hline
10 to 50 & 68 & 45 & 30 \\ 
\hline
51 to 80 & 36 & 26 & 16 \\ 
\hline
81 to 100 & 17 & 10 & 6 \\ 
\hline
101 to 120 & 15 & 13 & 11 \\ 
\hline
121 to 140 & 10 & 7 & 7 \\ 
\hline
141 to 160 & 12 & 11 & 10 \\ 
\hline
161 to 200 & 12 & 10 & 10 \\ 
\hline
201 to 220 & 5 & 2 & 2 \\ 
\hline
221 to 250 & 5 & 3 & 3 \\ 
\hline
251 to 270 & 1 & 1 & 1 \\ 
\hline
271 to 300 & 3 & 2 & 2 \\ 
\hline
301 to 380 & 1 & 1 & 1 \\ 
\hline
All & 200 & 131 & 99\\
\hline
\end{tabular}
\newline 
\caption{Open-Ephyra answer extraction results} 
\label{table: Open-Ephyra answer extraction results} 
\end {center}
\end{table}

Regarding Table \ref{table: Just.Ask answer extraction} we can see that the ratio CAE Success/AS Success increases with the increasing of extracted answers. Performing a more detalied analysis the results, when the number of extracted candidate answers is low (between 1 and 10) if it happens to contain the correct answer (that is, the CAE stage was successful), then the AS stage chooses it as final answer. When the number of candidate answers is between 41 and 60 or above 100, if the CAE stage is successful the AS stage also tends do be successful. The remaining results do not show any particular relationship between the CAE and AS stages. 

As for Table \ref{table: Open-Ephyra answer extraction results} like Just.Ask, we can see that the ratio CAE Success/AS Success increases with the increasing of extracted answers. Performing a more detalied analysis the results, when the number of candidate answers is between 161 and 380, if the CAE stage is successful the AS stage is also successful. When the number of candidate answers is between 101 and 160, if the CAE stage is successful the AS stage also tends do be successful. The remaining results do not show any particular relationship between the CAE and AS stages. 

When comparing both tables two main considerations come to mind. First, it can be clearly seen that Open-Ephyra extracts a larger number of answers (up to 380 answers) when compared to Just.Ask (which extracts only up to 230). Second and most important, Open-Ephyra outperforms Just.Ask both in the CAE step by extracting 17 more candidate answers and in the AS step by selecting 6 more final answers.  This fact seems to corroborate the validity of \emph{Hypothesis 1} and also allows to draw a new conclusion: \emph{Conclusion 2 - Just.Ask answer selection approach can be improved by using the answer selection approach of Open-Ephyra}.

The better performance of Open-Ephyra CAE step seems to indicate that \emph{Hypothesis 1} is a valid hypothesis. To confirm this ``validity'' the extraction methods that Open-Ephyra used for achieving the results presented in the ``CAE Success'' column in Table \ref{table: Open-Ephyra answer extraction results}, were analyzed. Table \ref{table: Open-Ephyra CAE results} presents the results of this analysis.

\begin{table} [h!t]
\begin{center}
\begin{tabular}{| c | c | c| c| c|} 
\hline 
\multicolumn{5}{|c|}{Open-Ephyra - Candidate Answer Extraction} \\ 
\hline 
Extracted Answers & \#Question & Pattern Approach & Answer Type Approach & Both Approaches\\ 
\hline 
0 & 15 & - & - & - \\ 
\hline
10 to 50 & 68 & 7 & 35 & 3\\ 
\hline
51 to 80 & 36 & 4 & 17 & 5 \\ 
\hline
81 to 100 & 17 & 2 & 7 & 1 \\ 
\hline
101 to 120 & 15 & 2 & 11 & 10 \\ 
\hline
121 to 140 & 10 & 1 & 1 & 5 \\ 
\hline
141 to 160 & 12 & 2 & 0 & 9 \\ 
\hline
161 to 200 & 12 & 1 & 1 & 8 \\ 
\hline
201 to 220 & 5 & 0 & 1 & 1 \\ 
\hline
221 to 250 & 5 & 0 & 0 & 3 \\ 
\hline
251 to 270 & 1 & 0 & 1 & 0 \\ 
\hline
271 to 300 & 3 & 0 & 0 & 2 \\ 
\hline
301 to 380 & 1 & 0 & 1 & 0 \\ 
\hline
All & 200 & 19 & 65 & 47\\
\hline
\end{tabular}
\newline 
\caption{Results of the different Open-Ephyra approaches in the CAE step } 
\label{table: Open-Ephyra CAE results} 
\end {center}
\end{table}  

The table confirms the results found in Section \ref{sec: Overall Impact of Open-Ephyra Answer Extraction Techniques}, i.e, the Answer Type approach is the approach that leads to the extraction of more candidate answers (65 candidate answers). Regarding the Pattern approach, when its solely used it leads to the extraction of 19 candidate answers and when its used in combination with the Answer Type approach it enables the system to extract 47 candidates answers. The use of the Pattern approach (both solely and in conjunction with the Answer Type approach) allows the system to extract 66 candidates, which practically the same amount of answers extract by the Answer Type approach. 

At this moment we can state that the use of the Pattern Approach contributes significantly to the results obtained by Open-Ephyra in the CAE step, and therefore we can say that \emph{Hypothesis 1} is a valid hypothesis. With this another conclusion can be made: \emph {Conclusion 3 -  Just.Ask performance can be improved by the utilization of the Pattern Type approach (solely and in conjunction with the Question Classification approach)}.

\section {Feasibility Study of Suggested Improvements}

Throughout the last section three improvement strategies for Just.Ask were discovered based on strategies used by Open-Ephyra. This section provides a high level assessment of ``how hard'' is to implement those improvements in Just.Ask. For the following discussing it must be kept in mind that both Open-Ephyra and Just.Ask are written in Java.

The three improvements that were suggested in the last section are:
\begin{itemize}
\item Improvement 1 - Just.Ask performance can be improved by the utilization of Reformulation, Interpretation and Term queries used by Open-Ephyra.
\item Improvement 2 - Just.Ask answer selection approach can be improved by using the answer selection approach of Open-Ephyra.
\item Improvement 3 - Just.Ask performance can be improved by the utilization of the Pattern Type approach (solely and in conjunction with the Question Classification approach).
\end {itemize}

\subsection {Improvement 1}

The implementation of Reformulation queries is not that hard and could be achieved exactly like Open-Ephyra implemented them. Three classes would be necessary for that: QuestionReformulation (the class that represents the question reformulation); Question Reformulator (the class that deals with the actual creation of question reformulations from a set of defined patterns) and QueryReformulatorGenerator (the class that knows how to make queries for formulations of a question).  The first two classes could easily copied and pasted from Open-Ephyra into the package where are the classes that deal with the interpretation of the user question (some modifications would have to be made in order for them to work properly). The QueryReformulatorGenerator could also be copied and paste from Open-Ephyra although some major modifications would have to be made so that the class could work using Just.Ask's QueryFormulator interface.

The implementation of Interpretation queries follows the same reasoning as Reformulation queries, so it will not be described. 

The implementation of Term queries also fallows the same reasoning as Reformulation queries, but deserves a special mention. In Open-Ephyra The class TermExtractor (which extract the terms of a question) uses a set of NLP tools for that purpose that Just.Ask do not possess. The class could be implemented in Just.Ask trough regular copy/paste, but the libraries corresponding to the NLP tools used by Open-Ephyra would have to copied and integrated as well.

\subsection {Improvement 2}

The implementations of the several filters that Open-Ephyra use to select the final answer can be implemented in Just.Ask using a ``copy/paste'' approach. The way how candidate answers in Just.Ask are represented would have to be changed to the expected representation type that those filters receive as argument. In its turn, the way how the filters return their output would have to be changed so that Just.Ask could continue to presents the results in the same manner.

\subsection {Improvement 3}

The Pattern Type approach is probably the easiest improvement to implement since, it can be achieved by a direct copy/paste with minor modifications in order for the class work properly in Just.Ask.

\section {Conclusions}

In this work were introduced three QA systems -– Just.Ask, Aranea and Open Ephyra. Just.Ask is an ontology driven system that uses and combines a lot of state or the art strategies. In Just.Ask questions are interpreted with the help of NLP tools and its classification is performed using rule- or machine learning based- techniques. Answers can be obtained using different information sources (as Google and Wikipedia) and different query formulation techniques. Correct answers are extracted using a named entity recognizer. 

Aranea is a system based on two different approaches, one searches answers in structured resources and the other uses data redundancy of unstructured information sources (like the Web) to do so. In the later, several types of queries are formulated to exploit the redundancy factor. Relevant passages are retrieved by using the previous queries, and n-grams are generated for these passages. Answer candidates then go through a filtering process in order to improve their quality, and finally, several heuristics are utilized to extracted answers from answer candidates.   

Open Ephyra is a system based on two different approaches for answer extraction. Pattern learning techniques are used to learn patterns about questions/answers, which facilitates the extraction of correct answers for patterns that are recognized. This is the most notable feature of Open Ephyra. Answer type extraction is used in the same fashion as Question Classification is utilized in Just.Ask, e.g., answers are extracted based on the type of answer that is expected. 

A preliminary evaluation of these three systems allowed to identify the most important characteristics to take into account for a proper evaluation of these systems. These characteristics are: the type of processed questions, the corpus utilized and the evaluation measures used. Based on these characteristics it was concluded that all systems would be evaluated regarding factoid questions and the best setting for doing so was composed by: gold-QA as the question corpus, the Web as answer corpus and Just.Ask's evaluation measures for evaluate the performance of the systems. Difficulties in adapting Aranea to this setting led to the decision of performing Aranea's evaluation apart form the evaluation of the other systems. This decision allowed that the evaluation of both Open-Ephyra and Just.Ask could be extended to definition type questions.

Regarding Aranea's evaluation, some problems occurred when trying to put Aranea working properly. These problems resulted in a degradation of the systems performance since the system only gave answers to a very short number of questions. For these questions both Just.Ask and Open-Ephyra obtained better performance relatively to Aranea.

The extrinsic and intrinsic evaluation of Just.Ask and Open-Ephyra identified Open-Ephyra as the best QA system evaluated in this work. Some ideas for improvement of Just.Ask  - new query types, new answer extraction approaches and new answer selection approaches - were also obtained in this process.Although these improvement ideas were determined for Just.Ask they can be applied to other QA systems that have similarities with Just.Ask. More specifically, QA systems that only use Keyword Type queries and QA systems that extract candidate answers with the Question Classification approach.

% \section {Future Work}

%Using the proposed setting an extrinsic and intrinsic evaluation will be performed for Just.Ask, Aranea and Open Ephyra
%systems regarding factoid type questions. The extrinsic evaluation will allow to see in overall which from these systems is %the best. The intrinsic evaluation will allow (hopefully) to identify for each task of the pipelining process of QA: the %techniques from other QA systems that can be used to improve that task and the impact of the proficiency of that task in %further steps of the pipelining process.
%Although, the conclusions of this study will be for only three QA systems, hopefully some conclusions will be attained that %can be applied in the context of other QA systems.  

%\subsection {Query Expansion in Other QA Systems}

%\subsection {Query Expansion Experiments in Just.Ask}

%\section {Evaluation}

\bibliographystyle{splncs}
\bibliography{refs/bibliography}

\end{document}